\documentstyle[12pt]{article}
\def\dz{\partial_z}
\def\cdz{\partial_{\bar{z}}}
\def\m{\mu_z{}^{\bar{z}}}
\def\cm{\mu_{\bar{z}}{}^z}
\def\cD{\cal D}
\def\e{\vec{e}}
\def\p{\vec{p}}
\def\s{\prime}
\def\tz{\tilde{z}}
\def\i{\mbox{i}}

\def\be{\begin{equation}}
\def\ee{\end{equation}}
\def\eh{\mbox{e}}
\begin{document}
\section*{}
\hspace*{\fill} LMU--TPW 94--6, IASSNS--HEP--94/83\\
\hspace*{\fill} hep--th/9410099\\[3ex]

\begin{center}
\Large\bf
Half--Differentials and Fermion Propagators
\end{center}
\vspace{-6ex}
\section*{}
\normalsize \rm
\begin{center}
{\bf Rainer Dick}\\ {\small \it School of Natural Sciences, Institute
for Advanced Study\\ Olden Lane, Princeton, NJ 08540, USA \\ {\rm and}\\
Department of Physics,
University of Munich\\
Theresienstr.\ 37, 80333 Munich,
Germany
}

\end{center}

\section*{}
{\bf Abstract}:
 From a geometric point of view, massless spinors in $3+1$ dimensions are
composed of primary fields of weights $(\frac{1}{2},0)$ and
$(0,\frac{1}{2})$, where the weights are defined with respect
to diffeomorphisms of a sphere in momentum space.
The Weyl equation thus appears as a consequence
of the transformation behavior of local sections
of half--canonical bundles under a change of charts.
As a consequence, it is possible to impose covariant constraints
on spinors of negative (positive) helicity in terms of (anti--)holomorphy
conditions. Furthermore, the identification with half--differentials
is employed to determine possible extensions
of fermion propagators compatible with Lorentz covariance.
\newpage
\section{Introduction}
The quest for a four--dimensional notion of analyticity and the related
problem to define four--dimensional analogues of two--dimensional conformal
field theories is a subject of much interest and was studied extensively in
recent years. A possibility which attracted particular attention
relies on the ideas of
quaternionic analyticity \cite{gt}, either in terms of Fueter analyticity
\cite{gj},
or in terms of a harmonic space approach \cite{gikos,gios,ego},
which is strongly connected to the twistor approach to space--time
\cite{p1,pr}.

On the other hand, attempts to transfer methods of 2D conformal field theory
directly to 4D conformal field theories
led
to the discovery of
structures reminiscent
of Zamolodchikov's $c$--theorem \cite{c1,jo}, and to new results
on correlation functions in 4D conformal field theories,
including in particular an extension of the central
charge of 2D conformal field theory to a triple of central
charges in 4 dimensions \cite{op,o2}. Of related interest are recent
results on quasi-primary fields in the $O(N)$ $\sigma$--model
for $2<d<4$ \cite{lr}.

In the present paper, I will follow a different approach to transfer
notions of 2D conformal field theory into $3+1$ dimensions.
I would like to point out that
left or right handed massless spinors in $3+1$
dimensions can be interpreted as half--differentials on spheres in
{\sl momentum space}. This implies the possibility to formulate
covariant phase space constraints on spinors of definite helicity
in terms of (anti--)meromorphy constraints (or (anti--)holomorphy
constraints outside finitely many poles).
More specifically, the entries of a spinor
of negative helicity appear as local representations $\psi(z,\bar{z},E)$
with respect to a conformal atlas, and transform under holomorphic
transformations according to
\begin{equation} \label{tra}
\psi^{\prime}(z^{\prime},\bar{z}^{\prime},E^{\prime})
=\psi(z,\bar{z},E)
\left(\frac{\partial z^{\prime}}{\partial z}\right)^{-\frac{1}{2}}
\end{equation}
Special cases of this transformation behavior imply the Weyl equation for
massless fermions.
Lorentz transformations induce via $SL(2,\mbox{\bf C})$
holomorphic transformations
of spheres in momentum space, and the resulting transformation behavior
of left handed spinors agrees with the equation above.
Therefore,
left handed spinors can be subjected to conditions
\begin{equation} \label{hol}
\frac{\partial \psi}{\partial\bar{z}}=0
\end{equation}
which are covariant under Lorentz transformations.
In this sense, the identification of spinors of definite helicity with
half--differentials
induces notions of 2D conformal field theory in $3+1$
dimensions\footnote{Somewhat sloppy,
I will refer to (\ref{hol}) as a holomorphy
constraint. The degree of the divisor of
a meromorphic $\lambda$--differential
on a surface of genus $g$ (i.e.\ the sum of orders of zeros minus the
sum of pole orders) equals $2\lambda (g-1)$, so spinors
satisfying (\ref{hol})
have at least one first order pole or two poles of order
$\frac{1}{2}$ on the unit sphere in momentum space.}.

To
clarify the notion of half--differentials it is useful to develop
a covariant primary field formalism not relying on conformal gauges
\cite{dick,nic}.
This will be
reviewed in section 2. This section serves to explain in particular
how to covariantize the factorized transformation behavior of primary
fields under 2D diffeomorphisms, which translates via our construction
into a factorized transformation behavior of helicity spinors under
Lorentz transformations.
The relation between spinors and half--differentials
is then explained in detail in section 3.
In section 4 the relation between half--differentials and Weyl spinors
is exploited
to
fix the structure of massless Lorentz covariant fermion
propagators.
As a result we will find that
the Lorentz covariant fermion
correlation in the free massless limit is
determined up to two functions $f_1^{}$ and
$f_2^{}$ which depend on single, but different arguments \cite{d1}:
\begin{equation}\label{result}
\langle\Psi(\p\,)\overline{\Psi}(\p^{\,\prime})\rangle=
\end{equation}
\[
\left(\begin{array}{cc}0&1\\0&0\end{array}\right)\otimes
\left(\begin{array}{cc}\bar{z}z^\prime & \bar{z}\\
z^\prime&1\end{array}\right)\langle\phi(\p\,)\phi^+(\p^{\,\prime})\rangle+
\left(\begin{array}{cc}0&0\\1&0\end{array}\right)\otimes
\left(\begin{array}{cc}1 & -\bar{z}^\prime\\
-z& z\bar{z}^\prime
\end{array}\right)\langle\psi(\p\,)\psi^+(\p^{\,\prime})\rangle
\]
\[
+
\left(\begin{array}{cc}1&0\\0&0\end{array}\right)\otimes
\left(\begin{array}{cc}\bar{z} & -\bar{z}\bar{z}^\prime\\
1&-\bar{z}^\prime
\end{array}\right)\langle\phi(\p\,)\psi^+(\p^{\,\prime})\rangle +
\left(\begin{array}{cc}0&0\\0&1\end{array}\right)\otimes
\left(\begin{array}{cc}z^\prime & 1\\
-zz^\prime&-z\end{array}\right)\langle\psi(\p\,)\phi^+(\p^{\,\prime})\rangle
\]
\begin{equation}\label{f1}
\langle\psi(\p_1^{})\psi^+(\p_2^{})\rangle=
\langle\phi(\p_2^{})\phi^+(\p_1^{})\rangle=
f_1^{}\!\left(\frac{|\p_1^{}|}{|\p_2^{}|}\right)
\frac{1+z_1^{}\bar{z}_2^{}}{\sqrt{|\p_1^{}||\p_2^{}|}}\,
\delta_{z\bar{z}}^{}(z_1^{}-z_2^{})
\end{equation}
\begin{equation}\label{f2}
\langle\psi(\p_1^{})\phi^+(\p_2^{})\rangle=
\overline{\langle\phi(\p_2^{})\psi^+(\p_1^{})\rangle}=
\frac{1}{z_1^{}-z_2^{}}\,
f_2^{}\!\left(|\p_1^{}||\p_2^{}|
\frac{(z_1^{}-z_2^{})(\bar{z}_1^{}-\bar{z}_2^{})}
{(1+z_1^{}\bar{z}_1^{})(1+z_2^{}\bar{z}_2^{})}\right)
\end{equation}
where $z(\p\,)$ is a stereographic coordinate in momentum space:
\[
z=\frac{p_+}{|\vec{p}\,|-p_3}
\]

The full derivation of these results will be given
in section 4.
Of special interest are the
$f_2^{}$--terms, since these terms
are the only terms which comply both with Lorentz
covariance and chiral symmetry breaking. Note the consistency of this
result: Since (\ref{result}) provides a
Lorentz covariant {\sl massless} propagator, those parts of it which
break chiral symmetry must also break translational
invariance. This is in agreement with Eq.\ (\ref{f2}), since the right
hand side of this equation cannot accomodate for a $\delta$--function
in external momenta. The $f_1^{}$--terms in turn preserve chiral symmetry:
They
do not contribute to a chiral condensate and anticommute with $\gamma_5^{}$.
Consistency of the result in this sector is expressed by the fact that these
terms contain
a $\delta$--function which restricts the correlator to parallel momenta.

Indeed, one motivation for the work reported in section 4
arose from low energy QCD:
Hadron spectroscopy and QCD sum rules provide evidence that
chiral symmetry remains broken in the low energy sector of QCD
even in the limit $m_q^{}\to 0$ \cite{ynd}.
Therefore, the
$f_2^{}$--terms
might shed new light on the problem of
chiral symmetry breaking in low--energy QCD.

Applications of methods of 2D field theory
also proved very useful in certain kinematical
regions of high energy QCD,
see \cite{lip,vv}.

For recent investigations of confinement and chiral symmetry breaking in the
framework of
supersymmetric gauge theories
see \cite{sw,is} and references therein.
Exploiting holomorphy
constraints on superpotentials and effective couplings
with respect to chiral superfields
and microscopic couplings yields a whole wealth of
non--perturbative results on the phases of these theories,
and in particular strong evidence for the appearance of massless
monopoles at certain points of the moduli space.
It is very remarkable, that even modest
perturbations of the superpotentials trigger monopole condensation, thus
yielding confinement via the dual Meissner effect \cite{sw,is}.

Besides chiral symmetry breaking, motivation for the present work came
also
from the observation that quantum group symmetric Heisenberg relations
break translational invariance by introducing an exponential
discretization in any direction, which however cannot be interpreted
naively as a lattice structure \cite{jul1}. Thus quantum groups may
provide a highly unusual and very attractive possibility to achieve
an immanent regularization of quantum field theory with enough
remnants of lattice structures to ensure finiteness, but without spoiling
Lorentz symmetry.

However, I should emphasize that the results of section 4 are independent
of any particular dynamics and also do not rely on any quantum group concepts.
Eq.\ (\ref{result}) essentially constitutes the result of a group theoretical
investigation of massless fermion propagators in the free limit by use
of the mapping between half--differentials and Weyl spinors described
in section 3.

\section{Covariant Primary Fields}
In two--dimensional field theories two apparently different formulations
of covariance existed in parallel for several years. On the one hand
two--dimensional field theories can be formulated covariantly in the
usual way employing tensor and spinor fields,
while on the other hand
it is known that in a conformal gauge primary fields can be employed
to ensure covariance with respect to the conformal remnant of the
diffeomorphism group \cite{bpz}.
This was puzzling, because there exist primary fields of half--integral
order on two--manifolds, and it was not clear in
what sense these could be considered as remnants
of tensor or spinor fields in a conformal
gauge\footnote{In this section, spinor refers to 2D spinors}.
Furthermore, it
was unclear how half--differentials should transform under non--conformal
transformations, or how they could be defined outside the realm of
conformal gauge fixing.
The puzzle was partially solved by the introduction of a
covariant definition of primary fields \cite{dick},
thus demonstrating
that primary fields yield factorized
representations of the full two--dimensional diffeomorphism group.
This work also included a demonstration of
isomorphy between tensor fields and covariant primary fields of integer
weight. However, the exact relation between spinors in two dimensions
and the covariant
half--differentials of \cite{dick} was given only recently in \cite{nic},
where the formalism was further developed and
applied to two--dimensional supergravity.

Initially primary fields $\Phi$ of conformal weight
$(\lambda,\bar{\lambda})$
on a two--manifold ${\cal M}$
are defined by their transformation behavior
under a holomorphic change of charts $z\to u(z)$ \cite{bpz}:
\begin{equation} \label{dpa}
\Phi(u,\bar{u}) = \Phi(z,\bar{z})\cdot \left({\frac{\partial
u}{\partial z}}\right)^{-\lambda} \cdot \left({\frac{\partial \bar{u}}{\partial
\bar{z}}}\right)^{-\bar{\lambda}}
\end{equation}
where I
employed the usual convention to denote the weight for the complex
conjugate sector of coordinates by $\bar{\lambda}$.

The scaling dimension of the field $\Phi$ is
$\Delta = \lambda+ \bar{\lambda}$
and the
spin\footnote{We distinguish between the spin $\sigma$ referring
to rotations induced by diffeomorphisms of ${\cal M}$ and the spin
$s$ referring to rotations of tangent frames.}
is $\sigma = \lambda - \bar{\lambda}$.
A cohomological investigation reveals that
the spin is restricted
to integer
or half--integer values, while no similar restriction is imposed on
the scaling dimension. We will demonstrate this in the more general setting
of covariant primary fields below.

The factorized transformation behavior makes primary fields particular
convenient for the formulation of two--dimensional field theories and
the investigation of short distance expansions.
However, this definition of primary fields
works only in a conformal gauge, i.e.\ in an atlas
with holomorphic transition functions. This causes no problem for integer
values of $\lambda$ and $\bar{\lambda}$, because the corresponding primary
fields might be considered as remnants of tensor fields in the conformal gauge.
However, such an interpretation is not possible for fractional conformal
weights. Furthermore, if the metric of the two--manifold $\cal M$ is
considered as a dynamical degree of freedom it is very inconvenient to
switch to a conformal gauge, because this implies that two degrees of freedom
of the metric corresponding to the Beltrami--parameters (see below)
are hidden in the holomorphic transition functions.
Therefore, in a conformal gauge
it is impossible to formulate the dynamics
of the metric in terms of local fields.

To avoid the restriction to conformal atlases
requires a generalization of equation (\ref{dpa})
to diffeomorphisms
$z \to u(z,{\bar{z}})$, i.e.\ we will define primary fields for arbitrary
atlases on smooth two--manifolds, thereby introducing a covariant definition
of half--differentials. Hence, in the sequel $z,w$ and $u$ will denote
complex local coordinates, but no holomorphy conditions on transformations
will be assumed any more.
To define covariant primary fields
it is convenient to switch to a Beltrami--parametrization
of the metric:
\begin{equation} \label{metric}
(ds)^2={\frac{2\,g_{z\bar{z}}}{1+\mu_{\bar{z}}{}^z\cdot\mu_z{}^{\bar{z}}}}
\cdot \left|dz+\mu_{\bar{z}}{}^z\cdot d\bar{z}\right|^2
\end{equation}
i.e.\ the Beltrami--parameters $\{\m,\cm\}$
specify the metric modulo scaling
transformations:
\begin{eqnarray} \label{bel1}
\cm &=& \frac{g_{z\bar{z}}^{}-
\sqrt{g_{z\bar{z}}^2-g_{zz}^{}g_{\bar{z}\bar{z}}^{}}}
{g_{zz}^{}}\\
{}&=& \frac{g_{\bar{z}\bar{z}}^{}}{g_{z\bar{z}}^{}+
\sqrt{g_{z\bar{z}}^2-g_{zz}^{}g_{\bar{z}\bar{z}}^{}}} =
\overline{\mu_z{}^{\bar{z}}}
\nonumber\\ \label{bel2}
\frac{g_{zz}^{}}{g_{z\bar{z}}^{}} &=& \frac{2\mu_z{}^{\bar{z}}}
{1+\mu_{\bar{z}}{}^z\mu_z{}^{\bar{z}}}
\end{eqnarray}
The Beltrami--parameters satisfy $\mu_{\bar{z}}{}^z\mu_z{}^{\bar{z}} < 1$
and have a subtle transformation behavior under reparametrizations
$z\to u(z,\bar{z})$ with
$\left|\partial_z u\right| >
\left| \partial_{\bar{z}} u\right| $:
\begin{equation} \label{trb}
\mu_{\bar{u}}{}^u = \frac{\mu_{\bar{z}}{}^z \cdot \partial_z u
- \partial_{\bar{z}} u}{\partial_{\bar{z}} \bar{u} - \mu_{\bar{z}}{}^z \cdot
\partial_z \bar{u}}
= \frac{\partial_{\bar{u}} z + \mu_{\bar{z}}{}^z \cdot \partial_{\bar{u}}
\bar{z}}{\partial_u z + \mu_{\bar{z}}{}^z \cdot \partial_u \bar{z}}
\end{equation}
This transformation law implies in particular
\[
\partial_{\bar{z}}-\cm \dz = (\cdz \bar{u} - \cm\dz \bar{u})
(\partial_{\bar{u}} - \mu_{\bar{u}}{}^u \partial_u) =
\frac{1}{\partial_{\bar{u}}\bar{z} - \mu_{\bar{u}}{}^u \partial_u\bar{z}}
(\partial_{\bar{u}} - \mu_{\bar{u}}{}^u \partial_u)
\]

This observation motivates the introduction of particular
non--holonomic bases of vector fields and differentials on two--manifolds
$\cal M$:
\begin{eqnarray} \label{pbs1}
{\cal D}_z &=& \partial_z - \mu_z{}^{\bar{z}} \cdot
\partial_{\bar{z}} \\*[1ex] \label{pbs2}
{\cal D}z &=& \frac{1}{1-\mu_{\bar{z}}{}^z \cdot \mu_z{}^{\bar{z}}}
\left(dz + \mu_{\bar{z}}{}^z \cdot d{\bar{z}}\right) \\[1ex] \label{pbs3}
\partial_z &=& \frac{1}{1-\mu_{\bar{z}}{}^z\cdot\mu_z{}^{\bar{z}}}
\left({\cal D}_z + \mu_z{}^{\bar{z}} \cdot {\cal D}_{\bar{z}}\right) \\*[1ex]
\label{pbs4}
dz &=& {\cal D}z-\mu_{\bar{z}}{}^z\cdot{\cal D}{\bar{z}}
\end{eqnarray}

These bases are distinguished by their factorized transformation
properties under diffeomorphisms:
\begin{equation} \label{tbs}
{\cal D}_u = \left({\cal D}_u z\right)\,{\cal D}_z \qquad {\cal D}u = {\cal
D}z\,{\cal D}_z u \qquad {\cal D}_z u = \left({\cal D}_u z\right)^{-1}
\end{equation}
thus allowing us to introduce
a consistent covariant definition of primary fields:\\[2ex]
{\bf Definition:}
A field $\Phi$ over a two--manifold is denoted as {\em primary} of weight
$(\lambda,\bar{\lambda})$
if its local representations $\Phi(z,\bar{z})$
transform under a change of coordinates $z,\bar{z} \to u,\bar{u}$ according to
\begin{equation}\label{defprim}
\Phi(u,\bar u) = \Phi(z,\bar z)\cdot \left({\cal D}_z u\right)^{-\lambda}
\cdot \left({\cal D}_{\bar z} {\bar u}\right)^{-\bar \lambda}
\end{equation}

\vspace{2ex}
In particular any tensor representation of the diffeomorphism group factorizes
into appropriate primary fields with integer weights upon expansion with
respect to the non--holonomic bases (\ref{pbs1},\ref{pbs2}),
but the crucial point is that
fractional weights can be defined as well
without conformal gauge fixing.

As we remarked before,
there is a restriction on the admissible values of the weight
$(\lambda,\bar \lambda)$: In a region of three intersecting patches
$U_I^{},U_J^{},U_K{}$ with coordinates
$z_I^{}, z_J^{}, z_K^{}, z_I^{} = f_{IJ}^{}(z_J^{},\bar{z}_J^{})$, etc.,
the product of transition functions for a roundtrip
$z_I^{} \to z_J^{} \to z_K^{} \to z_I^{}$ must yield the identity:
\begin{equation} \label{tcond}
({\cD}_{z_K}f_{IK}^{})^\lambda({\cD}_{z_J}f_{KJ}^{})^\lambda({\cD}_{z_I}
f_{JI}^{})^\lambda
({\cD}_{\bar{z}_K}\bar{f}_{IK}^{})^{\bar\lambda}
({\cD}_{\bar{z}_J}\bar{f}_{KJ}^{})^{\bar\lambda}
({\cD}_{\bar{z}_I}\bar{f}_{JI}^{})^{\bar\lambda} = 1
\end{equation}
For integer weights this condition is automatically fulfilled due to
$f_{KI}^{} = f_{KJ}^{}\circ f_{JI}^{}$ and eq.\ (\ref{tbs}).
However, if $\Delta=\frac{r}{s},\sigma=\frac{p}{q}$
are the representations of $\Delta$ and $\sigma$ in terms
of integers without common divisors,
and if $q\neq 1$,
then
it is a non--trivial problem to fix the $q$--fold ambiguity in the
definition of the transition functions
$
\left({\cal D}_{z_I}f_{JI}^{}\right)^{\lambda}
\cdot \left({\cal D}_{{\bar z}_{I}} {\bar f}_{JI}^{}\right)^{\bar \lambda}
$
in the intersections of all patches in such a manner that the condition
(\ref{tcond}) is fulfilled. To elaborate this further, we split the
transition functions into modulus and phase according to
\[
{\cal D}_{z_J}f_{IJ}^{} = R_{IJ}^{}\exp (i\phi_{IJ}^{})
\]
If we now stick to the convention to choose $R_{IJ}^{\frac{1}{s}}$
positive real
in any intersection $U_I^{} \cap U_J^{}$, then (\ref{tcond}) reduces to
\begin{equation} \label{scond}
\exp (i\sigma\phi_{IK}^{})\cdot \exp (i\sigma\phi_{KJ}^{})\cdot
\exp (i\sigma\phi_{JI}^{}) = 1
\end{equation}
and this defines the choice of phases as a sheaf--cohomological problem:\\
To clarify this define
\begin{equation} \label{ds}
S_{IJK}^{} \equiv
\exp (i\sigma\phi_{IK}^{})\cdot \exp (i\sigma\phi_{KJ}^{})\cdot
\exp (i\sigma\phi_{JI}^{})
\end{equation}
which is an element of $Z_q$.
Consider the sheaf ${\cal M}\times Z_q$ with base manifold $\cal M$
and stalk $Z_q$. An $n$--cochain is a completely antisymmetric
functional of intersections of $n+1$ patches with values
in $Z_q$:
\begin{eqnarray*}
c(U_{I(0)}^{}\cap U_{I(1)}^{}\cap\ldots\cap U_{I(n)}^{})
&=&c_{I(0)I(1)\ldots I(n)}^{}=c_{I(1)I(0)\ldots I(n)}^{-1}\in Z_q
\\
c(\emptyset)&=&1
\end{eqnarray*}
Then there are coboundary operators $\delta_n$ in the
pre--sheaf related to the cover $\{U_I^{}\}$ mapping $n$--cochains
to $(n+1)$--cochains:\\
\begin{eqnarray*}
(\delta_0^{} c)_{IJ}^{} &=& \frac{c_I^{}}{c_J^{}} \\
(\delta_1^{} c)_{IJK}^{} &=& c_{IJ}^{}\frac{1}{c_{IK}^{}}c_{JK}^{} \\
(\delta_2^{} c)_{IJKL}^{} &=& c_{IJK}^{}\frac{1}{c_{IJL}^{}}c_{IKL}^{}
\frac{1}{c_{JKL}^{}}
\end{eqnarray*}
and we have
\[ \delta_{n+1}^{}\delta_n^{}=1\]
Then $S$ as defined in (\ref{ds}) is a closed 2--cochain:
$\delta_2^{}S = 1$.
Unfortunately this does not imply exactness of $S$, because
the phase factors $\exp(i\sigma\phi_{IJ})$ generically do not
satisfy $x^q=1$. On the other hand exactness is what we are seeking,
because in this case
we would have
\begin{eqnarray*}
S_{IJK}^{} &\equiv&
\exp (i\sigma\phi_{IK}^{})\cdot \exp (i\sigma\phi_{KJ}^{})\cdot
\exp (i\sigma\phi_{JI}^{})\\
{}&=& (\delta_1^{} \theta)_{IJK}^{} = \theta_{IJ}^{}\theta_{JK}^{}
\theta_{KI}^{}
\end{eqnarray*}
for some 1--cochain $\theta$ in ${\cal M}\times Z_q$
and we could rescale the phase factors
$\exp (i\sigma\phi_{IJ}^{}) \to \exp (i\sigma\phi_{IJ}^{})\theta_{JI}^{}$
such that the condition (\ref{scond}) could be fulfilled.
Therefore, we may admit only those values for the
denominator $q$ of the spin, which correspond
to a trivial cohomology group $H^2({\cal M},Z_q)$.
However, it is a classical result on two--manifolds that this
cohomology group equals $\emptyset$ for every ${\cal M}$ if and only if
$q=1$ or $q=2$ \cite{hs}.
Hence,
the spin of primary fields over two--manifolds
is restricted to integral or half--integral values.
This implies in particular that the fractional values of conformal
weights
appearing in the conformal grids of minimal models must be combined
into the weights $(\lambda , \bar{\lambda })$
of primary fields such that $\sigma$
is half--integer or integer.
This
rule seems also justified empirically, because it is
in agreement with
the weights appearing in explicit realizations of
minimal models.

Let us now take a closer look at the correspondence between tensors and spinors
on the one hand and primary fields on the other hand:

As remarked before, the isomorphy between tensors and primary fields
of integer weight is given by expansion with respect to the
anholonomic basis (\ref{pbs1},\ref{pbs2}).
More specifically, we denote
a tensor $T$ with $m$ covariant and $n$ contravariant
indices as a tensor of covariance $\langle m,n\rangle$.
Upon expansion with respect to the primary basis (\ref{pbs1},\ref{pbs2})
a tensor of covariance $\langle m,n\rangle$ decays into
$2^{m+n}$ primary fields according to the reduction formula
\[
\langle m,n\rangle = \sum_{i=0}^m\sum_{j=0}^n {m\choose i}{n\choose j}
(i-j,m-n-i+j)
\]
For a tensor $T$ of covariance $\langle 2,0\rangle$ the transformation
to the related primary fields ${\cal T}$ is given by:
\begin{eqnarray*}
{\cal T}_{zz}^{} &=& T_{zz}^{}-\m (T_{\bar{z}z}^{}+T_{z\bar{z}}^{})
+\m\m T_{\bar{z}\bar{z}}^{}\\
{\cal T}_{z\bar{z}}^{} &=& T_{z\bar{z}}^{}-\m T_{\bar{z}\bar{z}}^{}
-\cm T_{zz}^{}
+\m\cm T_{\bar{z}z}^{}\\
T_{zz}^{} &=& \frac{1}{(1-\m\cm )^2}
\left({\cal T}_{zz}^{}+\m ({\cal T}_{\bar{z}z}^{}+{\cal T}_{z\bar{z}}^{})
+\m\m {\cal T}_{\bar{z}\bar{z}}^{}\right)\\
T_{z\bar{z}}^{} &=& \frac{1}{(1-\m\cm )^2}
\left({\cal T}_{z\bar{z}}^{}+\m {\cal T}_{\bar{z}\bar{z}}^{}+
\cm {\cal T}_{zz}^{}
+\m\cm {\cal T}_{\bar{z}z}^{}\right)
\end{eqnarray*}
and the conjugate formulas.
A special case is the metric, where the formulas above yield ${\cal G}_{zz}=
{\cal G}_{\bar{z}\bar{z}}=0$ and
\[
{\cal G}_{z\bar{z}}^{}=g_{z\bar{z}}^{} \frac{(1-\m\cm )^2}{1+\m\cm }
\]
In the primary field formalism we presented here the metric is represented
by a real primary field ${\cal G}_{z\bar{z}}$ and a complex Beltrami parameter
$\m$:
\[(ds)^2 = 2{\cal G}_{z\bar{z}}{\cal D}z\cdot {\cal D}\bar{z}\]
Note however that we might choose as well any other
definite symmetric tensor
of covariance $\langle 2,0\rangle$, construct the corresponding
Beltrami--parameters and derive another covariant primary field
formalism in exactly the same manner.


The relation between two--dimensional spinors
and covariant half--differentials
has been clarified by employing an appropriate zweibein
formalism \cite{nic}.
Therefore, consider complex orthogonal bases in the tangent
frames:
\[
\vec{e}_{\zeta}^{}=\frac{1}{2}({\e}_1^{}-i{\e}_2^{})
\]
\[
\eta_{\zeta\zeta}^{}=0,\qquad \eta_{\zeta\bar{\zeta}}^{} = \frac{1}{2}
\]
We stick to the convention that greek indices transform under the
symmetry group of the tangent bundle, while latin indices refer to
transformations under diffeomorphisms.
Remember that in the complex orthogonal
bases rotations in the tangent bundle are diagonal:
\[
\Lambda (\phi) = \left(\begin{array}{cc} \mbox{e}^{i\phi} & 0\\
0 & \mbox{e}^{-i\phi}\end{array}
\right)
\]
For spinors we choose a Weyl basis
$\gamma_1^{}=\sigma_1^{},\gamma_2^{}=\sigma_2^{}$
such that the spinor representation of $SO(2)$ is diagonal as well:
\[
S(\phi) = \left(\begin{array}{cc} \exp(\frac{i}{2}\phi) & 0\\
0 & \exp(-\frac{i}{2}\phi)\end{array}
\right)
\]
In the zweibein formalism the Beltrami parameters appear as ratios
of zweibein components: Insertion of
\[ g_{zz}^{}= e_{z}{}^{\zeta}e_{z}{}^{\bar{\zeta}} \qquad
g_{z\bar{z}}^{}=
\frac{1}{2}(e_{z}{}^{\zeta}e_{\bar{z}}{}^{\bar{\zeta}}+
e_{z}{}^{\bar{\zeta}}e_{\bar{z}}{}^{\zeta})
\]
into
(\ref{bel1}) yields
\begin{equation}\label{zb}
e_{\bar{z}}{}^{\zeta} = \cm e_{z}{}^{\zeta}
\end{equation}
Therefore, the primary zweibein which transforms like a primary field
of weight (1,0) under diffeomorphisms is
\[
\varepsilon_z{}^{\zeta} = e_{z}{}^{\zeta} (1-\m\cm )
\]
Equation (\ref{zb}) implies for the inverse zweibein
\begin{equation}\label{izb}
e^{\bar{z}}{}_{\zeta}= -\m e^{z}{}_{\zeta}
\end{equation}
and therefore {\em the diagonal components of the inverse zweibein are
primary fields of weight} $(-1,0)$ {\em and} $(0,-1)$ {\em respectively}:
\[
\varepsilon^z{}_{\zeta} = e^{z}{}_{\zeta} = \frac{1}{\varepsilon_z{}^{\zeta}}
\]
Thus $e^{z}{}_{\zeta}$ transforms under factorized representations
both under the diffeomorphism group and the tangent space rotations.
Therefore the transformation behavior of fractional powers of
$e^{z}{}_{\zeta}$ is well behaved.
More specifically, $(e^{z}{}_{\zeta})^{-\lambda}
(e^{\bar{z}}{}_{\bar{\zeta}})^{-\bar{\lambda}}$ is a primary field
of weight $(\lambda ,\bar{\lambda})$ under diffeomorphisms and a field of
spin
$s = \bar{\lambda}-\lambda$ under tangent space rotations,
and we know by
(\ref{scond}) that $s$ is restricted to integer and half--integer
values.
In particular, the sought for isomorphy between covariant half--differentials
$\psi_{\sqrt{z}}$ of weight $(\frac{1}{2},0)$ and chiral Weyl spinors
$\psi_{\sqrt{\zeta}}$ is \cite{nic}
\begin{equation}\label{iso}
\psi_{\sqrt{z}}^{}\sqrt{e^{z}{}_{\zeta}} = \psi_{\sqrt{\zeta}}^{}
\end{equation}

Having established equivalence between tensors and spinors on the one
hand and covariant primary fields on the other hand, it is also desirable
to develop a covariant primary differential calculus.
Therefore, we introduce a covariant primary derivative $D_z$ which maps
primary fields of weight $(\lambda ,\bar\lambda )$ and spin $s$ into
primary fields of the same spin and weight $(\lambda +1,\bar\lambda )$:
\begin{equation}\label{covder}
D_z^{} \Phi = {\cD}_z^{} \Phi -\lambda \Gamma^z{}_{zz}\Phi
-\bar{\lambda}\Gamma^{\bar{z}}{}_{\bar{z}z}\Phi -is\Omega_z^{}
\Phi
\end{equation}
Covariance of this construction with respect to diffeomorphisms $z \to u(z,
\bar{z})$ and rotations ${\e}_{\zeta}\to {\e}_{\zeta}\exp(-i\phi)$ implies
\begin{eqnarray}\label{conn1}
\Gamma^u{}_{uu}&=&({\cD}_z^{}u)^{-1}\Gamma^z{}_{zz}-({\cD}_z^{}u)^{-2}
{\cD}_z^{}{\cD}_z^{}u\\ \label{conn2}
\Gamma^{\bar{u}}{}_{\bar{u}u}&=&
({\cD}_z^{}u)^{-1}\Gamma^{\bar{z}}{}_{\bar{z}z}-({\cD}_z^{}u)^{-1}
({\cD}_{\bar{z}}^{}\bar{u})^{-1}{\cD}_z^{}{\cD}_{\bar{z}}^{}\bar{u}\\
\label{conn3}
\Omega_u^{} &=&
({\cD}_z^{}u)^{-1}(\Omega_z^{}+{\cD}_z^{}\phi)
\end{eqnarray}
In applications of this formalism in two--dimensional field theory
there frequently appear the anholonomy coefficients of the primary
bases (\ref{pbs1},\ref{pbs2}), because these coefficients
automatically appear as connection coefficients, if conformally gauge fixed
actions like the Ising model or the bosonic string are covariantized
in this formalism \cite{dick}:
\begin{eqnarray*}
[{\cD}_z^{},{\cD}_{\bar z}^{}] &=& C^{\bar{z}}{}_{\bar{z}z}{\cD}_{\bar z}
- C^z{}_{z\bar{z}}{\cD}_z^{}\\
d{\cD}z &=& C^z{}_{z\bar{z}}{\cD}z\wedge {\cD}\bar{z}\\
C^{\bar{z}}{}_{\bar{z}z} &=& \frac{1}{1-\cm\m }({\cD}_{\bar z}^{}\m
-\m {\cD}_z^{}\cm )
\end{eqnarray*}
The commutator of the covariant primary derivatives is then
\begin{equation}\label{comm}
[D_z^{},D_{\bar{z}}^{}]\Phi = (C^{\bar{z}}{}_{\bar{z}z} -
\Gamma^{\bar{z}}{}_{\bar{z}z})D_{\bar{z}}^{}\Phi
-
(C^z{}_{z\bar{z}} -
\Gamma^z{}_{z\bar{z}})D_z^{}\Phi - \lambda{\cal R}_{z\bar{z}}^{}\Phi
+\bar{\lambda}{\cal R}_{\bar{z}z}^{}\Phi - is{\cal F}_{z\bar{z}}^{}\Phi
\end{equation}
with curvature and field strength
\begin{eqnarray*}
{\cal R}_{z\bar{z}}^{} &=& {\cD}_z^{}\Gamma^z{}_{z\bar{z}}
- {\cD}_{\bar z}^{}\Gamma^z{}_{zz} - C^{\bar z}{}_{\bar{z}z}
\Gamma^z{}_{z\bar{z}} + C^z{}_{z\bar{z}}\Gamma^z{}_{zz}\\
{\cal F}_{z\bar{z}}^{} &=& {\cD}_z^{}\Omega_{\bar{z}}^{}-{\cD}_{\bar{z}}
\Omega_z^{}-C^{\bar z}{}_{\bar{z}z}\Omega_{\bar{z}}^{}+
C^z_{z\bar{z}}\Omega_z^{}
\end{eqnarray*}
Thus curvatures consist of
primary fields
of weight (1,1) in this formalism. However, due to the absence of
second order terms in the connection coefficients, $\cal R$ is not
a mere translation of the ordinary curvature tensor into the primary basis.

Similar to the tensor formalism one may impose constraints on the
connection:
The requirement of invariance of the metric under parallel translations
implies
\begin{equation}\label{metcon}
\Gamma^z{}_{zz} = {\cD}_z^{}\ln({\cal G}_{z\bar{z}}^{})-
\Gamma^{\bar z}{}_{\bar{z}z}
\end{equation}
while the requirement of vanishing torsion implies
\begin{equation}\label{torcon}
\Gamma^{\bar z}{}_{\bar{z}z} = C^{\bar z}{}_{\bar{z}z}
\end{equation}
The consistency of the torsion constraint
with (\ref{conn2}) follows easily from the transformation
behavior of the Lie bracket.

On the other hand, one may also impose a zweibein postulate:
\begin{eqnarray*}
D_z^{}e^z{}_{\zeta} &=& 0\\
D_z^{}e^{\bar z}{}_{\bar{\zeta}} &=& 0
\end{eqnarray*}
implying
\begin{eqnarray*}
i\Omega_z^{} &=& \Gamma^z{}_{zz} + {\cD}_z^{}\ln (e^z{}_{\zeta})\\
{}&=& - \Gamma^{\bar z}{}_{\bar{z}z} - {\cD}_z^{}
\ln (e^{\bar z}{}_{\bar{\zeta}})
\end{eqnarray*}
The zweibein postulate implies in particular invariance of the
metric under parallel translations (\ref{metcon}).

\section{Massless Fermions and Half--Differentials}
Half--differentials turn out to appear not only in
two--dimensional field theories,
but also in 3+1 dimensions, because
space-time spinors of definite helicity define half--differentials
on a sphere in momentum space and vice versa. To explain this, it
is convenient to employ the Weyl representation for Dirac matrices,
and to parametrize the unit sphere in momentum space in terms of
stereographic coordinates:
\begin{equation}\label{zdef1}
z=\frac{p_+}{|\vec{p}\,|-p_3}
\end{equation}
\begin{equation} \label{zdef2}
\tilde{z}=-\frac{1}{z}
\end{equation}
For later use we also give the inversion formulas:
\be\label{inv1}
p_1^{}=|\p\,|\frac{z+\bar{z}}{z\bar{z}+1}
\ee
\be \label{inv2}
p_2^{}=\i|\p\,|\frac{\bar{z}-z}{z\bar{z}+1}
\ee
\be \label{inv3}
p_3^{}=|\p\,|\frac{z\bar{z}-1}{z\bar{z}+1}
\ee
The metric on the unit sphere reads in terms of these coordinates
\[
ds^2=\frac{4dzd\bar{z}}{(1+z\bar{z})^2}
\]
and hence the zweibein of the previous section in this case
is given by
\[
e_{z}{}^{\zeta}=\frac{2}{1+z\bar{z}}\]
\[
e_{\bar{z}}{}^{\zeta}=0
\]
Therefore, primary derivatives reduce to ${\cal D}_z^{}=\partial_z^{}$,
while
the covariant derivative of a primary field of weight
$(\lambda,\bar{\lambda})$ and spin $s$ is given by
\[
D_z^{}\Phi=\partial_z^{}\Phi+(2\lambda+s)\frac{\bar{z}}{1+z\bar{z}}\Phi
\]

The relation between the local representations $\psi(z,\bar{z},|\p\,|)$ and
$\psi(\tilde{z},\bar{\tilde{z}},|\p\,|)$ of a primary field
of weight $(\frac{1}{2},0)$ is according to (\ref{defprim})
\begin{equation}\label{weyl}
\psi(\tilde{z},\bar{\tilde{z}},|\p\,|)= -z\psi(z,\bar{z},|\p\,|)
\end{equation}
where the sign ambiguity has been resolved in such a way to avoid minus signs
in the expressions for Weyl spinors and Dirac spinors below.
Insertion of the definition of $z$ demonstrates that this
is exactly
the Weyl equation for a massless spinor with opposite signs of
chirality and energy:
\[
(|\p\,|+\p\cdot\vec{\sigma})
\left(\begin{array}{c}\psi(z,\bar{z},|\p\,|)\\
\psi(\tilde{z},\bar{\tilde{z}},|\p\,|)\end{array}\right)=0
\]
Similarly, the relation between local representations of a primary field
of weight $(0,\frac{1}{2})$
\begin{equation}\label{antiweyl}
\phi(\tilde{z},\bar{\tilde{z}},|\p\,|)= \bar{z}\phi(z,\bar{z},|\p\,|)
\end{equation}
is the Weyl equation for a massless spinor of equal signs of energy
and chirality:
\[
(|\p\,|-\p\cdot\vec{\sigma})\left(\begin{array}{c}
\phi(\tilde{z},\bar{\tilde{z}},|\p\,|)\\ \phi(z,\bar{z},|\p\,|)
\end{array}\right)=0
\]
Half--differentials thus yield spinor
bases:
\begin{eqnarray} \label{span1}
\Psi^{(++)}(\p\,)&=&\left(\begin{array}{c}1\\0
\end{array}\right)
\otimes \left(\begin{array}{c}\bar{z}\\1\end{array}\right)
\phi(z,\bar{z},|\p\,|)\\ \label{span2}
\Psi^{(-+)}(\p\,)&=&\left(\begin{array}{c}0\\1
\end{array}\right)
\otimes \left(\begin{array}{c}\bar{z}\\1\end{array}\right)
\phi(z,\bar{z},|\p\,|)\\ \label{span3}
\Psi^{(+-)}(\p\,)&=&\left(\begin{array}{c}1\\0
\end{array}\right)
\otimes \left(\begin{array}{c}1\\-z\end{array}\right)
\psi(z,\bar{z},|\p\,|)\\ \label{span4}
\Psi^{(--)}(\p\,)&=&\left(\begin{array}{c}0\\1
\end{array}\right)
\otimes \left(\begin{array}{c}1\\-z\end{array}\right)\psi(z,\bar{z},|\p\,|)
\end{eqnarray}
Here the first superscript denotes chirality, while the
second superscript indicates the
helicity. The spinors with equal signs of helicity and chirality are the
positive energy solutions.

To prove covariance
of this construction under Lorentz transformations, we first identify the
transformation behavior of $z(\p\,)$ to prove then from (\ref{defprim})
that the two local representations of a half--differential transform like
the components of a Weyl spinor:

Under parity or time reversal $z(\p\,)$ goes to
$z(-\p\,)=-\bar{z}(\p\,)^{-1}$
and thus half--differentials of weight $(\frac{1}{2},0)$ become
half--differentials of weight $(0,\frac{1}{2})$ and vice versa.

Under proper orthochronous Lorentz transformations $\Lambda(\omega)
=\exp(\frac{1}{2}\omega^{\mu\nu}L_{\mu\nu}^{})$,
with $\omega$ the usual set of rotation and boost parameters,
$z(\p\,)$ goes to
\begin{equation}\label{zlor1}
z^{\prime}=z(\p^{\,\prime})=\overline{U}\circ z(\p\,)=
\frac{\bar{a}z+\bar{b}}{\bar{c}z+\bar{d}}
\end{equation}
if $E=|\p\,|$, and to
\begin{equation}\label{zlor2}
z^{\prime}=z(\p^{\,\prime})=U^{-1T}\circ z(\p\,)=
\frac{dz-c}{a-bz}
\end{equation}
if $E=-|\p\,|$.\\
Here $U$ is the positive chirality spin representation of $\Lambda$:
\[
U(\omega)=\exp(\frac{1}{2}\omega^{\mu\nu}\sigma_{\mu\nu}^{})
=\left(\begin{array}{cc} a & b\\ c & d\end{array}\right)\in
SL(2,\mbox{\bf C})
\]
\vspace*{1ex}
{\sl Proof of
the transformation law (\ref{zlor1}):}

The principle of the proof will be employed repeatedly in the sequel,
and works as follows: In order to prove transformation properties
under the proper orthochronous Lorentz group, we first check that the
proposed transformation behavior has the correct composition properties
under subsequent Lorentz transformations. In order to complete the proof
it is then sufficient to verify the proposed transformation properties
for rotations around two different axes and boosts in a particular
direction, because these transformations provide a generating set
for the full proper orthochronous Lorentz group, as will be explained
in Eq.\ (\ref{decom}).

First we observe that $\overline{U}\circ z=\overline{V}\circ z$ for all $z$
if and only if $U=\pm V$, and
\begin{equation}\label{connect}
\overline{U}_2^{}\circ\overline{U}_1^{}\circ z
=\overline{U}_3^{}\circ z
\end{equation}
for all $z$ if and only if $U_3^{}=\pm
U_2^{}\cdot
U_1^{}$.
Hence the transformation law (\ref{zlor1}) provides a projective
representation of $SL(2,\mbox{\bf C})$ or a true representation of
the proper orthochronous Lorentz group $\cal L_+^{\uparrow}$.

However, every proper orthochronous
Lorentz transformation can be decomposed into
subsequent transformations consisting of boosts in $\e_3^{}$--directions and
rotations around $\e_1^{}$--axes and $\e_3^{}$--axes, in the following way:
Every proper orthochronous Lorentz transformation can be written as a pure
rotation followed by a pure boost $B(\vec{u}\,)\cdot R(\vec{\phi})$
(see e.g.\ \cite{sz}). This can be decomposed further according to
\begin{equation}\label{decom}
B(\vec{u}\,)\cdot R(\vec{\phi}\,)=R(-\alpha_u^{}\e_3^{})\cdot
R(-\beta_u^{}\e_1^{})\cdot B(u\e_3^{})\cdot R(\beta_u^{}\e_1^{})\cdot
R((\alpha_u^{}+\gamma_{\phi}^{})\e_3^{})\cdot R(\beta_{\phi}^{}\e_1^{})\cdot
R(\alpha_{\phi}^{}\e_3^{})
\end{equation}
where
$\alpha_{\phi}^{}$, $\beta_{\phi}^{}$ and
$\gamma_{\phi}^{}$ are the Euler angles of $R(\vec{\phi})$, and
\[
\cos(\beta_u^{})=\e_3^{}\cdot\e_u^{}
\]
\[
\cos(\alpha_u^{})\sin(\beta_u^{})=-\e_2^{}\cdot\e_u^{}
\]

As a consequence of (\ref{connect},\ref{decom})
it is sufficient to verify (\ref{zlor1}) for
the set of generating transformations
in order to prove
that $z(\p\,)$ as defined in (\ref{zdef1}) really satisfies (\ref{zlor1})
for $E=|\p\,|$.

First consider a rotation around $\e_3^{}$: The corresponding
$SL(2,\mbox{\bf C})$--matrix is
\[
U(\phi\e_3^{})=\left(\begin{array}{cc} \exp(\frac{i}{2}\phi) & 0\\
0 & \exp(-\frac{i}{2}\phi)\end{array}
\right)
\]
while $p_+^{\s}=p_+^{}\mbox{e}^{-\i\phi}$ and hence
$z(\p\,^{\s})=z(\p\,)\mbox{e}^{-\i\phi}$.

For a rotation around $\e_1^{}$ the corresponding
$SL(2,\mbox{\bf C})$--matrix is
\[
U(\phi\e_1^{})=
\left(\begin{array}{cc} \cos(\frac{\phi}{2}) & \i\sin(\frac{\phi}{2})\\
\i\sin(\frac{\phi}{2}) & \cos(\frac{\phi}{2})\end{array}
\right)
\]
while $z(\p\,^{\s})$ is
\[
z(\p\,^{\s})=\frac{p_1^{}+\i p_2^{}\cos(\phi)+\i p_3^{}\sin(\phi)}
{|\p\,|-p_3^{}\cos(\phi)+p_2^{}\sin(\phi)}=
\frac{z\cos(\frac{\phi}{2})-\i\sin(\frac{\phi}{2})}{\cos(\frac{\phi}{2})
-\i z\sin(\frac{\phi}{2})}
\]

Finally, for a boost along $\e_3^{}$
with boost parameter $u$ we have
\[
U(u\e_3^{})=\left(\begin{array}{cc} \exp(-\frac{u}{2}) & 0\\
0 & \exp(\frac{u}{2})\end{array}
\right)
\]
while $z(\p\,^{\s})=\mbox{e}^{-u}z(\p\,)$.
The transformation law (\ref{zlor2}) is proved in the same
way. In the last step one only has to take care that
$|\p\,^{\s}|=|\p\,|\cosh(u)+p_3^{}\sinh(u)$
and $p_3^{\s}=p_3^{}\cosh(u)+|\p\,|\sinh(u)$ for $E=-|\p\,|$\quad$\bf \Box$
\vspace*{1ex}

To conclude the demonstration that half--differentials on
the unit sphere in
momentum space are Weyl spinors it remains to show that the
two local representations of a half--differential transform
like a Weyl spinor:

First assume $E=|\p\,|$:
A half--differential $\phi$ of weight $(0,\frac{1}{2})$
then transforms according to (\ref{defprim}) into
\be\label{traphi}
\phi^\s(z^\s,\bar{z}^\s,|\p^{\,\s}|)=(c\bar{z}+d)\phi(z,\bar{z},|\p\,|)
\ee
while a half-differential $\psi$ of weight $(\frac{1}{2},0)$
transforms according to
\be\label{trapsi}
\psi^\s(z^\s,\bar{z}^\s,|\p^{\,\s}|)=(\bar{c}z+\bar{d})\psi(z,\bar{z},|\p\,|)
\ee
However, due to (\ref{weyl},\ref{antiweyl}) this is equivalent to
\[
\left(\begin{array}{c}\phi^\s(\tz^\s,\bar{\tz}^\s,|\p^{\,\s}|)\\
\phi^\s(z^\s,\bar{z}^\s,|\p^{\,\s}|)\end{array}\right)=U\cdot
\left(\begin{array}{c}\phi(\tz,\bar{\tz},|\p\,|)\\
\phi(z,\bar{z},|\p\,|)\end{array}\right)
\]
\[
\left(\begin{array}{c}\psi^\s(z^\s,\bar{z}^\s,|\p^{\,\s}|)\\
\psi^\s(\tz^\s,\bar{\tz}^\s,|\p^{\,\s}|)\end{array}\right)
=U^{-1\dagger}\cdot
\left(\begin{array}{c}\psi(z,\bar{z},|\p\,|)\\
\psi(\tz,\bar{\tz},|\p\,|)\end{array}\right)
\]

Thus a half--differential of weight $(0,\frac{1}{2})$ is equivalent to a
spin--$\{\frac{1}{2},0\}$--representation of the proper orthochronous
Lorentz group $\cal L_+^{\uparrow}$,
while a half--differential of
weight $(\frac{1}{2},0)$ is equivalent to a
spin--$\{ 0,\frac{1}{2}\}$--representation of the proper orthochronous
Lorentz group if $E=|\p\,|$.

On the other hand, if $E=-|\p\,|$, then this corresponds to
$U\leftrightarrow U^{-1\dagger}$ in the equations above,
and the assignment
of the half--differentials $\phi$ and $\psi$
to representations of $\cal L_+^{\uparrow}$ is changed.

This concludes our demonstration that half--differentials
yield Weyl spinors of definite helicity in Minkowski space
and vice versa.

\section{Propagators}
As an application of the results of the last section we now
exploit the factorized transformation behavior of the half--differentials
under Lorentz transformations to determine the general structure
of massless fermion propagators compatible with Lorentz covariance.

The mode expansion of a
massless spinor contains positive and negative frequency
contributions:
\[
\Psi(x)=\frac{1}{\sqrt{2\pi}^3}\int\frac{d^3\p}{2|\p\,|}\left(\Psi_+^{}(\p\,)
\exp(\mbox{i}p\cdot x)+
\Psi_-^{}(\p\,)
\exp(-\mbox{i}p\cdot x)\right)
\]
The components $\Psi_{\pm}^{}(\p\,)$ have expansions
on helicity eigenstates, which can be expressed in terms of
half--differentials employing the results of the previous section:
\begin{equation}\label{expand}
\Psi(\p\,)=\left(\begin{array}{c}1\\0
\end{array}\right)
\otimes \left(\begin{array}{c}\bar{z}\\1\end{array}\right)
\phi(z,\bar{z},|\p\,|)+
\left(\begin{array}{c}0\\1
\end{array}\right)
\otimes \left(\begin{array}{c}1\\-z\end{array}\right)\psi(z,\bar{z},|\p\,|)
\end{equation}
The
expansions of
both the positive and negative energy contributions
contain only spinors with the same signs of chirality and helicity
since $\Psi^{c,h}(-\p\,)=\Psi^{c,-h}(\p\,)$. This is the reason for
the helicity $=$ chirality rule for massless fermions. As a consequence
of this reflection in the negative energy case, the half--differentials
in (\ref{expand}) transform according to (\ref{traphi},\ref{trapsi}),
irrespective of the sign of energy.

Eq.\ (\ref{expand})
yields representations of the corresponding correlation functions
in terms of primary fields:
\begin{equation}\label{prop}
\langle\Psi(\p\,)\overline{\Psi}(\p^{\,\prime})\rangle=
\end{equation}
\[
\left(\begin{array}{cc}0&1\\0&0\end{array}\right)\otimes
\left(\begin{array}{cc}\bar{z}z^\prime & \bar{z}\\
z^\prime&1\end{array}\right)\langle\phi(\p\,)\phi^+(\p^{\,\prime})\rangle+
\left(\begin{array}{cc}0&0\\1&0\end{array}\right)\otimes
\left(\begin{array}{cc}1 & -\bar{z}^\prime\\
-z& z\bar{z}^\prime
\end{array}\right)\langle\psi(\p\,)\psi^+(\p^{\,\prime})\rangle
\]
\[
+
\left(\begin{array}{cc}1&0\\0&0\end{array}\right)\otimes
\left(\begin{array}{cc}\bar{z} & -\bar{z}\bar{z}^\prime\\
1&-\bar{z}^\prime
\end{array}\right)\langle\phi(\p\,)\psi^+(\p^{\,\prime})\rangle +
\left(\begin{array}{cc}0&0\\0&1\end{array}\right)\otimes
\left(\begin{array}{cc}z^\prime & 1\\
-zz^\prime&-z\end{array}\right)\langle\psi(\p\,)\phi^+(\p^{\,\prime})\rangle
\]
Therefore, the 2--point functions on the right hand side
transform under a factorized
representation of the Lorentz group.
This makes this representation very convenient for the investigation
of all correlations $\langle\Psi(p)\overline{\Psi}(p^\prime)\rangle$
which comply with Lorentz covariance.

This is accomplished in the following way: The transformation behavior
of the 2--point functions on the right hand side of Eq.\ (\ref{prop})
is governed by (\ref{zlor1}) and by (\ref{traphi})
or (\ref{trapsi}) respectively.
Therefore, the determination of the general Lorentz covariant form
of the correlators $\langle\Psi(\p\,)\overline{\Psi}(\p^{\,\prime})\rangle$
is equivalent to the determination of the general form of correlators
of half-differentials complying with their respective transformation
properties.
Similar to the reasoning
employed in the proof of (\ref{zlor1},\ref{zlor2}) it is sufficient
to consider rotations $R(\phi\e_1^{})$ and $R(\phi\e_3^{})$, and
a boost $B(u\e_3^{})$ and solve the covariance conditions for
these transformations in order to ensure covariance with respect to
the full proper orthochronous Lorentz group, since the conformal factors
in (\ref{traphi},\ref{trapsi}) compose consistently under Lorentz
transformations.
Then invariance under parity {\bf P} or time reversal {\bf T} will
impose relations between the different 2--point functions and
automatically ensure invariance under charge conjugation.
This will uniquely fix the propagator up to two functions
$f_1^{}(|\p\,|/|\p\,^{\s}|)$
and $f_2^{}(-\frac{1}{2}p\cdot p^{\s})$.\\[1ex]
{\sl Determination of the correlation function}
$F_1^{}(z_1^{},z_2^{},\bar{z}_1^{},\bar{z}_2^{},|\p_1^{}|,|\p_2^{}|)=
\langle\psi(\p\,)\psi^+(\p^{\,\prime})\rangle$:

Eq.\ (\ref{trapsi}) implies invariance of $F_1^{}$
under rotations $R(\phi\e_3^{})$, and hence
\be\label{rot3}
F_1^{}=F_1^{}(z_1^{}\bar{z}_1^{},z_2^{}\bar{z}_2^{},z_1^{}\bar{z}_2^{},
|\p_1^{}|,|\p_2^{}|)
\ee
The other generating transformations yield more involved conditions on
$F_1^{}$:
Under rotations $R(\phi\e_1^{})$ $F_1^{}$ should transform according to
\be\label{conrot1}
F_1^{}\left(\frac{z_1^{}\bar{z}_1^{}\cos^2\left(\frac{\phi}{2}\right)
+\sin^2\left(\frac{\phi}{2}\right)+\i\cos\left(\frac{\phi}{2}\right)
\sin\left(\frac{\phi}{2}\right)(z_1^{}-\bar{z}_1^{})}
{z_1^{}\bar{z}_1^{}\sin^2\left(\frac{\phi}{2}\right)
+\cos^2\left(\frac{\phi}{2}\right)-\i\cos\left(\frac{\phi}{2}\right)
\sin\left(\frac{\phi}{2}\right)(z_1^{}-\bar{z}_1^{})},\right.
\ee
\[
\frac{z_2^{}\bar{z}_2^{}\cos^2\left(\frac{\phi}{2}\right)
+\sin^2\left(\frac{\phi}{2}\right)+\i\cos\left(\frac{\phi}{2}\right)
\sin\left(\frac{\phi}{2}\right)(z_2^{}-\bar{z}_2^{})}
{z_2^{}\bar{z}_2^{}\sin^2\left(\frac{\phi}{2}\right)
+\cos^2\left(\frac{\phi}{2}\right)-\i\cos\left(\frac{\phi}{2}\right)
\sin\left(\frac{\phi}{2}\right)(z_2^{}-\bar{z}_2^{})},
\]
\[
\left.
\frac{z_1^{}\bar{z}_2^{}\cos^2\left(\frac{\phi}{2}\right)
+\sin^2\left(\frac{\phi}{2}\right)+\i\cos\left(\frac{\phi}{2}\right)
\sin\left(\frac{\phi}{2}\right)(z_1^{}-\bar{z}_2^{})}
{z_1^{}\bar{z}_2^{}\sin^2\left(\frac{\phi}{2}\right)
+\cos^2\left(\frac{\phi}{2}\right)-\i\cos\left(\frac{\phi}{2}\right)
\sin\left(\frac{\phi}{2}\right)(z_1^{}-\bar{z}_2^{})},|\p_1^{}|,|\p_2^{}|
\right)=
\]
\[
[z_1^{}\bar{z}_2^{}\sin^2\left(\frac{\phi}{2}\right)
+\cos^2\left(\frac{\phi}{2}\right)-\i\cos\left(\frac{\phi}{2}\right)
\sin\left(\frac{\phi}{2}\right)(z_1^{}-\bar{z}_2^{})]
F_1^{}(z_1^{}\bar{z}_1^{},z_2^{}\bar{z}_2^{},z_1^{}\bar{z}_2^{},
|\p_1^{}|,|\p_2^{}|)
\]
When I solved this equation, I did it in terms of the differential equation
following from first order in $\phi$, checking then that the general solution
of that equation also solves the global condition. However, here is a slightly
more convenient solution: One may recognize that a
particular solution to (\ref{conrot1}) is given by
\[
F_1^{}=\frac{1}{1+z_1^{}\bar{z}_2^{}}
\]
This in turn implies that $F_1^{}$ may differ from the particular solution
at most by a factor $G_1^{}$
which is invariant under rotations both around $\e_1^{}$
and $\e_3^{}$--axes. However, since the
conformal factors in (\ref{conrot1}) compose
consistently under subsequent Lorentz transformations, this implies
invariance of
$G_1^{}$ under the full rotation group.
Therefore, the general solution of (\ref{rot3}) and (\ref{conrot1}) is
\be\label{rot13}
F_1^{}=\frac{1}{1+z_1^{}\bar{z}_2^{}}
G_1^{}\left(\frac{(z_1^{}-z_2^{})(\bar{z}_1^{}-\bar{z}_2^{})}
{(1+z_1^{}\bar{z}_1^{})(1+z_2^{}\bar{z}_2^{})},
|\p_1^{}|,|\p_2^{}|\right)
\ee
where the first argument is $p_1^{}\cdot p_2^{}$ up to normalization.

What remains to be checked are the boost properties of $F_1^{}$, and
by the reasoning employed in the proof of Eq.\ (\ref{zlor1}) it is sufficient
to
check boosts along $\e_3^{}$:\\
While the behavior of the parameters $(z,\bar{z},|\p\,|)$ under
rotations is completely specified by (\ref{zlor1}),
for a boost $B(u\e_3^{})$ we also have to specify the behavior of
$|\p\,|$:
\begin{eqnarray*}
z^\prime &=& \exp(- u)z\\
|\p\,^\prime|&=&\frac{|\p\,|}{z\bar{z}+1}\left(\exp(-u)z\bar{z}+
\exp(u)\right)
\end{eqnarray*}
Covariance of $F_1^{}$ then requires
\[
G_1^{}\left(\frac{(z_1^{}-z_2^{})(\bar{z}_1^{}-\bar{z}_2^{})}
{(\eh^u +\eh^{-u}z_1^{}\bar{z}_1^{})(\eh^u +\eh^{-u}z_2^{}\bar{z}_2^{})},
|\p_1^{}|\frac{\eh^u +\eh^{-u}z_1^{}\bar{z}_1^{}}{1+z_1^{}\bar{z}_1^{}},
|\p_2^{}|\frac{\eh^u +\eh^{-u}z_2^{}\bar{z}_2^{}}{1+z_2^{}\bar{z}_2^{}}\right)
\]
\be\label{conb3}
=
\frac{\eh^u +\eh^{-u}z_1^{}\bar{z}_2^{}}{1+z_1^{}\bar{z}_2^{}}
G_1^{}\left(\frac{(z_1^{}-z_2^{})(\bar{z}_1^{}-\bar{z}_2^{})}
{(1+z_1^{}\bar{z}_1^{})(1+z_2^{}\bar{z}_2^{})},
|\p_1^{}|,|\p_2^{}|\right)
\ee
The limit of large boost parameter shows that this equation can be solved
if and only if $z_1^{}=z_2^{}$, and hence $G_1^{}$ must contain a
2-dimensional $\delta$--function:
\[
G_1^{}=(1+z_1^{}\bar{z}_2^{})^2 \delta_{z\bar{z}}^{}(z_1^{}-z_2^{})
H_1^{}(|\p_1^{}|,|\p_2^{}|)
\]
Note that this result complies with Eq.\ (\ref{rot3}) since the 2-dimensional
$\delta$--function can be written in a fancy way:
\[
\delta_{z\bar{z}}^{}(z_1^{}-z_2^{})=-2z_1^{}\bar{z}_2^{}
\delta(z_1^{}\bar{z}_1^{}-z_2^{}\bar{z}_2^{})
\delta(z_1^{}\bar{z}_2^{}-z_2^{}\bar{z}_1^{})
\]
The covariance condition under boosts then assumes the following form:
\[
H_1^{}\left(|\p_1^{}|\frac{\eh^u +\eh^{-u}z\bar{z}}{1+z\bar{z}},
|\p_2^{}|\frac{\eh^u +\eh^{-u}z\bar{z}}{1+z\bar{z}}\right)
=
\frac{1+z\bar{z}}{\eh^u +\eh^{-u}z\bar{z}}
H_1^{}\left(
|\p_1^{}|,|\p_2^{}|\right)
\]
with general solution
\[
H_1^{}=f_1^{}\!\left(\frac{|\p_1^{}|}{|\p_2^{}|}\right)
\frac{1}{\sqrt{|\p_1^{}||\p_2^{}|}}
\]
Lorentz covariance thus fixes
the $(\frac{1}{2},0)\otimes (0,\frac{1}{2})$--differential
$\langle\psi(\p_1^{})\psi^+(\p_2^{})\rangle$ up to a function
$f_1^{}(|\p_1^{}|/|\p_2^{}|)$:
\begin{equation}\label{f11}
\langle\psi(\p_1^{})\psi^+(\p_2^{})\rangle=
f_1^{}\!\left(\frac{|\p_1^{}|}{|\p_2^{}|}\right)
\frac{1+z_1^{}\bar{z}_2^{}}{\sqrt{|\p_1^{}||\p_2^{}|}}\,
\delta_{z\bar{z}}^{}(z_1^{}-z_2^{})
\end{equation}
$\bf \Box$\\[1ex]
{\sl Determination of the correlation function}
$F_2^{}(z_1^{},z_2^{},\bar{z}_1^{},\bar{z}_2^{},|\p_1^{}|,|\p_2^{}|)=
\langle\psi(\p\,)\phi^+(\p^{\,\prime})\rangle$:\\
According to (\ref{trapsi}) and (\ref{traphi}) the covariance conditions
for rotations are for $R(\phi\e_3^{})$:
\be\label{2rot3}
F_2^{}(\eh^{-\i\phi}z_1^{},\eh^{-\i\phi}z_2^{},
\eh^{\i\phi}\bar{z}_1^{},\eh^{\i\phi}\bar{z}_2^{},|\p_1^{}|,|\p_2^{}|)=
\eh^{\i\phi}
F_2^{}(z_1^{},z_2^{},\bar{z}_1^{},\bar{z}_2^{},|\p_1^{}|,|\p_2^{}|)
\ee
and for $R(\phi\e_1^{})$:
\be\label{2rot1}
F_2^{}(z_1^{\s},z_2^{\s},\bar{z}_1^{\s},\bar{z}_2^{\s},|\p_1^{}|,|\p_2^{}|)
=
\ee
\[
\left(\cos\left(\frac{\phi}{2}\right)-\i z_1^{}\sin\left(\frac{\phi}{2}\right)
\right)
\left(\cos\left(\frac{\phi}{2}\right)-\i z_2^{}\sin\left(\frac{\phi}{2}\right)
\right)
F_2^{}(z_1^{},z_2^{},\bar{z}_1^{},\bar{z}_2^{},|\p_1^{}|,|\p_2^{}|)
\]
with
\[
z^{\s}=\frac{z\cos\left(\frac{\phi}{2}\right)-\i\sin\left(\frac{\phi}{2}\right)}
{\cos\left(\frac{\phi}{2}\right)-\i z\sin\left(\frac{\phi}{2}\right)}
\]
A special solution to these equations is given by
$(z_1^{}-z_2^{})^{-1}$, and hence the general solution is
\be\label{solvrot}
F_2^{}=\frac{1}{z_1^{}-{z}_2^{}}
G_2^{}\left(\frac{(z_1^{}-z_2^{})(\bar{z}_1^{}-\bar{z}_2^{})}
{(1+z_1^{}\bar{z}_1^{})(1+z_2^{}\bar{z}_2^{})},
|\p_1^{}|,|\p_2^{}|\right)
\ee
Covariance with respect to boosts then requires
\[
G_2^{}\left(\frac{(z_1^{}-z_2^{})(\bar{z}_1^{}-\bar{z}_2^{})}
{(\eh^u +\eh^{-u}z_1^{}\bar{z}_1^{})(\eh^u +\eh^{-u}z_2^{}\bar{z}_2^{})},
|\p_1^{}|\frac{\eh^u +\eh^{-u}z_1^{}\bar{z}_1^{}}{1+z_1^{}\bar{z}_1^{}},
|\p_2^{}|\frac{\eh^u +\eh^{-u}z_2^{}\bar{z}_2^{}}{1+z_2^{}\bar{z}_2^{}}\right)
\]
\be\label{2conb3}
=
G_2^{}\left(\frac{(z_1^{}-z_2^{})(\bar{z}_1^{}-\bar{z}_2^{})}
{(1+z_1^{}\bar{z}_1^{})(1+z_2^{}\bar{z}_2^{})},
|\p_1^{}|,|\p_2^{}|\right)
\ee
so the dependence of the left hand side on the boost parameter must
cancel identically, implying
\[
G_2^{}=
G_2^{}\left(\frac{(z_1^{}-z_2^{})(\bar{z}_1^{}-\bar{z}_2^{})}
{(1+z_1^{}\bar{z}_1^{})(1+z_2^{}\bar{z}_2^{})}
|\p_1^{}||\p_2^{}|\right)
\]
Therefore, the
$(\frac{1}{2},0)\otimes(\frac{1}{2},0)$--differential
$\langle\psi(\p_1^{})\phi^+(\p_2^{})\rangle$ takes the form
\begin{equation}\label{f22}
\langle\psi(\p_1^{})\phi^+(\p_2^{})\rangle=
\frac{1}{z_1^{}-z_2^{}}\,
f_2^{}\!\left(|\p_1^{}||\p_2^{}|
\frac{(z_1^{}-z_2^{})(\bar{z}_1^{}-\bar{z}_2^{})}
{(1+z_1^{}\bar{z}_1^{})(1+z_2^{}\bar{z}_2^{})}\right)
\end{equation}

Invariance under {\bf P} or {\bf T} then fixes the
remaining 2--point functions
\begin{eqnarray*}
\langle\psi(\p_1^{})\phi^+(\p_2^{})\rangle=
\overline{\langle\phi(\p_2^{})\psi^+(\p_1^{})\rangle}\\
\langle\psi(\p_1^{})\psi^+(\p_2^{})\rangle=
\langle\phi(\p_2^{})\phi^+(\p_1^{})\rangle
\end{eqnarray*}
thus establishing the result we were seeking.

As expected, Lorentz symmetry alone complies with a bilocal propagator
in momentum space, and it
restricts the chiral symmetry preserving
parts (\ref{f11}) to parallel momenta.
On the other hand, chiral symmetry breaking terms
must account for breaking of translational invariance. This is in agreement
with (\ref{f22}), because these correlation functions cannot accomodate for
$\delta$--functions preserving the direction of momentum.

The unperturbed result for the on--shell correlation
\[
\langle\psi(\p\,)\overline{\psi}(\p^{\,\prime})\rangle =
-2p\cdot\gamma|\p\,|\delta(\p-\p^{\,\prime})
\]
is recovered from Eqs.\ (\ref{prop},\ref{f11},\ref{f22}) for
$f_1^{}(x)=\delta(x-1)$, $f_2^{}=0$.

Off--shell extensions of (\ref{prop})
can be inferred from the requirement to yield the same
propagator in configuration space:
\begin{eqnarray}
S(x,x^{\prime})&=&\frac{\Theta(t-t^{\prime})}{(2\pi)^3}
\int\frac{d^3\p}{2|\p\,|}
\int\frac{d^3\p^{\,\prime}}{2|\p^{\,\prime}|}\exp(\mbox{i}p\cdot x)
\mbox{i}\langle\Psi(\p\,)\overline{\Psi}(\p^{\,\prime})\rangle
\exp(-\mbox{i}p^{\prime}\cdot x^{\prime})\nonumber\\
{}&-&
\frac{\Theta(t^{\prime}-t)}{(2\pi)^3}
\int\frac{d^3\p}{2|\p\,|}
\int\frac{d^3\p^{\,\prime}}{2|\p^{\,\prime}|}\exp(-\mbox{i}p\cdot x)
\mbox{i}\langle\Psi(\p\,)\overline{\Psi}(\p^{\,\prime})\rangle
\exp(\mbox{i}p^{\prime}\cdot x^{\prime})\nonumber\\
{}&=&\frac{1}{(2\pi)^4}\int d^4p
\int d^4p^{\prime}\exp(\mbox{i}p\cdot x)
S(p,p^{\prime})\exp(-\mbox{i}p^{\prime}\cdot x^{\prime}) \label{offprop}
\end{eqnarray}
thus fixing the structure up to the 2 functions $f_1^{}$, $f_2^{}$.
Insertion of the unperturbed on--shell correlation $f_1^{}(x)=\delta(x)$,
$f_2^{}=0$ yields the free Feynman propagator, of course.
However, it
is tempting to ask how QCD might account for the
chiral symmetry breaking $f_2^{}$--terms from a dynamical
point of view. Furthermore, low energy QCD should also imply a modification
of the $f_1^{}$--terms from the standard result, since confinement
seems hardly compatible with momentum conservation on the level
of single quark propagators.\\[2ex]
{\bf Acknowledgement:}
I would like to thank Hermann Nicolai and Julius Wess for
helpful discussions at various stages of this work. Support by the DFG is
gratefully acknowledged.

\end{document}